\newcommand{\be}{\begin{eqnarray}}
\newcommand{\ee}{\end{eqnarray}}
\newcommand{\nn}{\nonumber}

\newcommand{\lb}[1]{\label{#1}}

\newcommand\q{\quad}

\def\g{\gamma}

\def\d{\delta}

\def\ve{\varepsilon}

\def\pa{\partial}

\def\sfrac#1#2{{\textstyle\frac{#1}{#2}}}

\documentclass[12pt]{article}

\usepackage{amsmath,amssymb}
\usepackage{graphicx}
\usepackage{cite}
\usepackage{bm}

\begin{document}

\renewcommand{\thefootnote}{\fnsymbol{footnote}}

\vskip 15mm

\begin{center}

{\Large $6D$ superconformal theory as the Theory of Everything.}

\vskip 3ex  

A.V. \textsc{Smilga}

\vskip 3ex

\textit{SUBATECH, Universit\'e de
Nantes,  4 rue Alfred Kastler, BP 20722, Nantes  44307, France
\footnote{On leave of absence from ITEP, Moscow, Russia.}}
\\
e-mail: \texttt{smilga@subatech.in2p3.fr}
\end{center}

\vskip 5ex

\begin{abstract}
We argue that the fundamental Theory of Everything is a conventional field theory defined in the  
{\it flat} 
multidimensional bulk. Our Universe should be obtained as a 3-brane classical solution 
in this theory. The renormalizability
of the fundamental theory implies that it involves higher derivatives (HD).
It should be supersymmetric 
(otherwise one cannot get rid of the huge induced cosmological term) and probably
conformal (otherwise one can hardly cope with the problem of ghosts) . We present arguments 
that in conformal HD theories  the ghosts (which are  inherent for HD theories) 
might be not so malignant. 
In particular, we present a  nontrivial QM HD model 
where ghosts are {\it absent} and the spectrum has a well defined ground state. 

The requirement of superconformal invariance restricts the dimension of the 
bulk to be $D \leq 6$. We suggest that the TOE  lives in six dimensions and 
enjoys the maximum ${\cal N} = (2,0)$ superconformal symmetry.
Unfortunately, no  renormalizable field theory with this symmetry is presently known.       
 We construct and discuss an ${\cal N} = (1,0)$  $6D$ supersymmetric gauge theory with 
four derivatives in the action. 
This theory involves  a dimensionless coupling
constant and is renormalizable. At the tree level, the theory enjoys conformal symmetry, 
but the latter is broken
by quantum anomaly. The sign of the $\beta$ function corresponds to the Landau zero situation.
\end{abstract}

\renewcommand{\thefootnote}{\arabic{footnote}}
\setcounter{footnote}0
\setcounter{page}{1}

\section{Motivation}

Arguably, the most burning unresolved problem of modern theoretical physics is the absense of a satisfactory quantum theory
of gravity. The {\it main} obstacle here is the geometric nature of gravity. Time is intertwined there with spatial
coordinates and the notion of universal flat time is absent.
 As a result, in constrast to  conventional field theory,
one cannot write the (functional) Schr\"odinger equation, 
define the Hilbert space with unitary evolution operator, etc. 

As a matter of fact, Einstein gravity (and any other theory where the 
metric is considered as a fundamental dynamic variable) has problems also at the classical level. The equations of motion
cannot be {\it always} formulated as Cauchy problem. This leads to breaking of causality for some exotic configurations
like G\"odel universes or wormholes 
\cite{noncause}.

Even though these configurations are not realized in our world at the macroscopic 
level, their existence presents conceptual difficulties.

The modern paradigm is that the 
fundamental Theory of Everything is a form of string theory. If this is true, 
gravity has the 
status of  effective theory and one is not allowed to blame it for inconsistencies. But string theory also does not 
provide a satisfactory answer  to all these troubling questions. Actually, they cannot
even be posed there: we understand more or less well what  string theory is 
 only at the perturbative
level (and even there we are not sure yet whether technical difficulties preventing one now to perform calculation of string
amplitudes beyond  two loops 
 can be efficiently resolved),  while its non-perturbative formulation is simply 
absent.

This has led us to suggest \cite{duhi} 
that the TOE is a {\it field theory} living in  {\it flat} higher-dimensional space. 
This higher-dimensional
theory should involve 3-brane classical solutions, which might be associated with our Universe
in the spirit of \cite{Rub}. The gravity is induced
there as an effective theory living on the brane. One can imagine a thin soap bubble. Its effective hamiltonian is
  \be
 \label{Hsoap} 
H^{\rm eff} \ =\ \sigma \int \sqrt{g}\,  d^2x\ ,
  \ee
 where $\sigma$ is the surface tension. The hamiltonian (\ref{Hsoap}) is geometric, but the fundamental theory 
of soap is not: it is formulated in flat $3D$ space and does not know anything about the metric, etc. Of course, 
the analogy is not exact because the effective
hamiltonian (\ref{Hsoap}) does not have an Einstein form  but looks rather as a cosmological term. 
The Einstein term and also
the terms involving higher powers of curvature appear as corrections, however.
 In the observable world, the cosmological term
is either zero or very small and one should think of a mechanism to get rid of it. One could succeed in that (if any) only
if the fundamental theory is supersymmetric. Indeed, only supersymmetry can provide for the exact calcellation of quantum
corrections to the energy density of the brane solution. 

If we want the fundamental higher-dimensional theory to be renormalizable, the canonical dimension of the lagrangian
should be greater than 4, i.e. it should involve higher derivatives. HD theories  are known to have a problem of 
ghosts, which in many cases break unitarity and/or causality of the theory. \cite{PU}  
\footnote {Physically, a ghost--ridden theory is simply a theory where the 
spectrum has no bottom and one cannot 
define what vacuum is.} 
However, a model study performed in Refs. 
\cite{duhi,bezduhov}
indicates that in some cases, namely, when the theory enjoys {\it exact} conformal invariance,  the ghosts are not 
so malignant, a well defined ground state (the vacuum) might exist and the theory might enjoy a unitary S-matrix.

We conclude that the TOE should be superconformal  theory. This restricts the number of  dimensions $D$ in the 
flat space-time
 where the theory is formulated by $D \leq 6$. Indeed, all superconformal algebras involving the 
super-Poincare algebra as a
subalgebra are classified \cite{sconf}. Their highest possible dimension is six, 
which allows for the minimal conformal superagebra
(1,0) and the extended chiral conformal superalgebra (2,0). 

Our hypothesis is that the TOE lives in six dimensions and
enjoys the highest possible supersymmetry (2,0).

Unfortunately, no field theory with this symmetry group is actually known now. 
The corresponding lagrangian is not  constructed, and only indirect
results concerning scaling behavior of certain operators have been obtained so far \cite{Sok}. In \cite{ISZ}, 
we derived (using the formalism of harmonic superspace (HSS) \cite{HSS})  
the lagrangian for the $6D$ gauge theory with unextended $(1,0)$ superconformal symmetry. This theory
is conformal at the classical level and renormalizable. However, it is not finite: the $\beta$ function does not
vanish there and  conformal symmetry  is broken at the quantum level by 
anomaly. In other words, this theory  cannot be regarded as a viable candidate for the TOE.  
Its study represents, however, a necessary step before the problem
of constructing and studying the $(2,0)$ theory could be tackled. 
 
 In the next section, we explain in more details {\it what} are the ghosts, why (if not dealt with) 
they make the theory sick, and also 
present a special QM HD model where the ghosts {\it are} tamed.  In sect. 3 we derive the lagrangian 
of our superconformal $6D$ theory
and calculate its beta function.
 The last section is devoted, as usual, to conclusions and speculations.

 \section{Ghost-free QM higher derivative model.}
\setcounter{equation}0

To understand the nature  of ghosts, one does not need  to study 
field theories. 
It is clearly seen in toy models with finite
number of degrees of freedom. Consider e.g. the lagrangian
 \be
\label{Om4}
{ L} \ =\ \frac 12 \ddot q^2 - \frac {\Omega^4}2  q^2  \ .
 \ee 
It is straightforward to see that four independent solutions to the corresponding classical equations of motion are 
$q_{1,2}(t) = e^{\pm i\Omega t}$, $q_3(t) = e^{-\Omega t}$ and $q_4(t) = e^{\Omega t}$. The exponentially rising solution
$q_4(t)$ displays instability of the classical vacuum $q=0$. The quantum hamiltonian of such a system is not hermitian
and the evolution operator $e^{-i\hat H t}$ is not unitary.

 This vacuum instability is characteristic for all {\it massive} HD field theories  --- the 
dispersive equation has complex solutions in this case for small enough momenta. 
But for intrinsically massless (conformal) field theories
the situation is different. Consider the lagrangian
   \be
\label{L4mix}
{     L} \ =\ \frac 12 (\ddot q  + \Omega^2 q )^2 -  \frac \alpha 4 q^4 - \frac {\beta}2 q^2 \dot q^2 \ .
 \ee
Its quadratic part can be obtained from the HD field theory lagrangian ${\cal L} = (1/2) \phi \Box^2 \phi$
involving massless scalar field, when restricting it on the modes with a definite momentum $\vec{k}\ \  
(\Omega^2 = \vec{k}^2)$.
 If neglecting the nonlinear terms in (\ref{L4mix}), the solutions of the classical equations of motion
$q(t) \sim e^{\pm i\Omega t}$ and $q(t) \sim te^{\pm i\Omega t}$ do not involve exponential instability, but include only
comparatively ``benign'' oscillatory solutions with linearly rising  amplitude. 

We showed in \cite{duhi} that, when nonlinear terms in Eq.(\ref{L4mix}) are included, an island of stability in the 
neighbourhood of the classical vacuum
\footnote{Usually, the term classical vacuum is reserved for the point in the configuration (or phase) space with 
minimal energy. For HD theories and in particular for the theory (\ref{L4mix}) the classical 
energy functional is not bounded
from below and by ``classical vacuum'' we  simply mean a stationary  solution to the classical equations of motion.} 
 \be
 \label{vac}
q = \dot q = \ddot q = q^{(3)} = 0
 \ee
exists in a certain range of the parameters $\alpha, \beta$. In other words, when initial conditions are chosen 
at the vicinity
of this point, the classical trajectories $q(t)$ do not grow, but display a decent oscillatory behaviour. 
This island  is surrounded by the sea of instability, however. For generic initial conditions, the trajectories 
become singular: $q(t)$ and its derivatives reach infinity in a finite time.

Such a singular behaviour of classical trajectories often means trouble also in the quantum case. A well-known example 
when it does is the problem of $3D$ motion in the potential
 \be
\label{potr2}
V(r) \ =\ - \frac \gamma {r^2}\ .
  \ee
The classical trajectories where the particle falls to the centre (reaches
 the singularity $r=0$ in a finite time) are abundant.
This occurs when $l > \sqrt{2m\gamma}$, where $l$ is the classical angular momentum.
And it is also well known that, if $m\gamma > 1/4$, the quantum problem is not very well defined: the eigenstates with arbitrary 
negative energies exist and the hamiltonian does not have a ground state. 

 The bottomlessness of the quantum hamiltonian is not, however, 
a necessary corollary of the fact that the classical problem
involves singular trajectories. In the problem (\ref{potr2}), the latter are present for all positive $\gamma$, 
but the quantum ground state disappears only when $\gamma$ exceeds the boundary value $1/(4m)$.  

 Our main observation here is that the system (\ref{L4mix}) 
exhibits a similar behaviour. 
If both $\alpha$ and $\beta$
are nonnegative (and at least one of them is nonzero), the quantum hamiltonian has a bottom and the quantum problem
is perfectly well defined even though some classical trajectories are singular.

\subsection{Free theory}
 
Before analyzing the full nonlinear system (\ref{L4mix}), let us study the dynamics of the truncated system
with the lagrangian $ L = (\ddot q + \Omega^2 q)^2/2$. As was observed in \cite{DM}, this system displays a singular
behavior.  It is instructive to consider first the lagrangian
  \be
 \label{om12}
 { L} \ =\ \frac 12 \left[  \ddot q^2 - (\Omega_1^2 + \Omega_2^2) \dot q^2 + \Omega_1^2 \Omega_2^2 q^2 \right]
  \ee
and look what happens in the limit $\Omega_1 \to \Omega_2$.  When $\Omega_1 > \Omega_2$, the spectrum of the theory (\ref{om12}) is 
\be
\label{spec12}
E_{nm} = \left(n+ \frac 12 \right) \Omega_1 - \left( m + \frac 12 \right) \Omega_2 
\ee
with nonnegative integer $n,m$. On the other hand, when $\Omega_1 = \Omega_2 = \Omega$, the spectrum is
 \be
 \label{spec}
E_n =n\Omega
\ee
with generic integer $n$. In both cases, the quantum hamiltonian has no ground state, but in the limit of equal
frequencies the number of degrees of freedom is apparently reduced in a remarkable way: instead of two quantum numbers $n,m$
(the presence of two quantum numbers is natural --- the phase space of the system (\ref{om12}) is 4--dimensional
having two pairs $(p_{1,2}, q_{1,2})$ of canonic variables), we are left with only one quantum number $n$.

This deficiency of the number of eigenstates compared to natural expectations
 would not surprise a mathematician.
 A generic
$2\times 2$ matrix has two different eigenvectors. But the Jordan cell 
$\left( \begin{array}{cc} 1 & 1 \\ 0 & 1 \end{array}
\right) $
has only {\it one} eigenvector $\propto \left( \begin{array}c 1 \\ 0 \end{array} \right) $. 
The statement is therefore that in the 
limit $\Omega_1 = \Omega_2$ our hamiltonian represents a kind of generalized Jordan cell.

 Actually, the ``lost'' degrees of freedom reinstall themselves when taking into account nontrivial 
{\it time dynamics}
of the degenerate system (\ref{om12}) with $\Omega_1 = \Omega_2$. The situation is rather similar 
to what has been unravelled back in the sixties when studying degenerate systems displaying 
``nonexponential decay'' behavior (see e.g. \cite{Peres}). We will not discuss  here nonstationary problem 
and concentrate on the 
{\it spectrum} of  this system.

To begin with, let us construct the canonical hamiltonian corresponding to the 
lagrangian (\ref{om12}). This can be done 
using the general Ostrogradsky formalism \cite{Ostr}
\footnote{See e.g. \cite{Ham} for its detailed pedagogical description.}. 
For a lagrangian like (\ref{om12}) involving $q, \dot q$, and $\ddot q$, it consists in introducing the new variable
$x = \dot q$ and writing the hamiltonian $H(q,x; p_q, p_x)$ in such a way that the classical Hamilton equations
of motion would coincide after excluding the variables $x, p_x, p_q$ with the equations of motion
 \be
\label{eqmot}
q^{(4)} + (\Omega_1^2 + \Omega_2^2) \ddot q + \Omega_1^2 \Omega_2^2 q \ =\ 0
 \ee
derived from the lagrangian (\ref{om12}). This hamiltonian has the following form
  \be
\label{Ham12}
H \ =\ p_q x + \frac {p_x^2}2 + \frac {(\Omega_1^2 + \Omega_2^2) x^2}2 - \frac {\Omega_1^2 \Omega_2^2 q^2}2 \ .
  \ee
For example, the equation $\partial H/\partial p_q = \dot q$ gives the constraint $x = \dot q$ , etc. 

When $\Omega_1 \neq \Omega_2$,
the quadratic hamiltonian (\ref{Ham12}) can be diagonalized by a certain canonical transformation
$x, q, p_x, p_q \ \to a_{1,2}, a^*_{1,2}$ \cite{DM,bezduhov}.  
We obtain 
  \be
\label{Hdiag}
H \ =\ \Omega_1 a_1^* a_1 - \Omega_2 a_2^* a_2 \ .
  \ee
The classical dynamics of this hamiltonian  is simply $a_1 \propto e^{-i\Omega_1 t}$, 
$a_2 \propto e^{i\Omega_2 t}$. Its quantization  gives the spectrum (\ref{spec12}).
The negative sign of the second term in (\ref{Hdiag}) implies the negative sign of the corresponding kinetic term,
which is usually interpreted as the presence of the ghost states (the states with negative norm) in the spectrum.
We prefer to keep the norm positive definite, with the creation and annihilation operators $a_{1,2}, a_{1,2}^\dagger$
(that correspond to the classical variables   $a_{1,2}, a_{1,2}^*$) satisfying the usual commutation relations
$
[a_1, a_1^\dagger] \ =\ [a_2, a_2^\dagger]\ =\ 1\ $.
However, irrespectively of whether the metric is kept positive definite or not and the world ``ghost'' is used or not, 
the
spectrum (\ref{spec12}) does not have a ground state and, though the spectral problem for the free hamiltonian
(\ref{Hdiag}) is perfectly well defined, the absence of the ground state leads to a trouble, the falling to the centre
phenomenon when switching on the interactions. 
\footnote{A characteristic feature of this phenomenon is that some classical trajectories reach singularity in a finite
time while the quantum spectrum  involves a {\it continuum} of states with arbitrary low energies
\cite{fc}.
In our case, the ``centre'' is not a particular point in the configuration (phase) space but rather
its boundary at infinity, but the physics is basically the same.}

We are interested, however, not in the system (\ref{om12}) as such, but rather in this system in the limit
 $\Omega_1 =\Omega_2$. As was mentioned, this limit is singular.
 The best way to see what happens is to write down the explicit expressions for the wave 
functions of the states (\ref{spec12}) and
explore their behaviour in the equal frequency limit. 
This can be done by  substituting the operators $-i \partial/\partial x,
\ -i \partial /\partial q$ for $p_x$ and $p_q$ in Eq. (\ref{Ham12}) and  searching for the solutions of the Schr\"odinger
equation in the form
  \be
 \label{AnsPsi}
\Psi(q,x)\ =\ e^{-i\Omega_1 \Omega_2 qx} \exp\left\{ - \frac \Delta 2 \left( x^2 + 
\Omega_1 \Omega_2 q^2 \right) \right\} \phi(q,x)\ ,
 \ee
where $\Delta = \Omega_1 - \Omega_2$. Then the operator acting on $\phi(q,x)$ is
 \be
\label{Opphi}
 \tilde H \ =\ - \frac 12 \frac {\partial^2}{\partial x^2} + \left( \Delta x + i\Omega_1 \Omega_2 q \right) 
\frac \partial {\partial x} - ix \frac \partial {\partial q} + \frac \Delta 2\ .
 \ee
It is convenient to introduce 
 \be
z = \Omega_1 q + ix\ ,\ \ \ \ \  \ \ 
u = \Omega_2 q - ix \ ,
 \ee
after which the operator (\ref{Opphi}) acquires the form
  \be
 \label{Hhol}
 \tilde H(z,u) \ =\ \frac 12 \left( \frac \partial {\partial z} - \frac \partial {\partial u} \right)^2 + 
\Omega_1 u \frac \partial {\partial u} - \Omega_2 z \frac \partial {\partial z} + \frac \Delta 2 \ .
  \ee 
The holomorphicity of $ \tilde H(z,u)$ means that its eigenstates are holomorphic functions $\phi(z,u)$. An obvious
eigenfunction with the eigenvalue $\Delta/2$  is $\phi(z,u) = $ const. Further, if assuming  $\phi$ to be the function of
only one holomorphic variable $u$ or $z$, the equation $\tilde H \phi = E \phi$ acquires the same form as for the equation
for the preexponential factor in the standard oscillator problem. Its solutions are Hermit polynomials,
 \be
\label{n0i0m}
\phi_n(u) &=& H_n(i\sqrt{\Omega_1} u ) \equiv H_n^+ ,\ \ \ \ \ \ E_n = \frac \Delta 2 + n\Omega_1 \ , \nonumber \\
\phi_m(z) &=& H_m(\sqrt{\Omega_2} z ) \equiv H_m^- ,\ \ \ \ \ \ E_m = \frac \Delta 2 - m \Omega_2 \ .   
  \ee
 
The solutions (\ref{n0i0m}) correspond to excitations of only one of the oscillators while another one is in its 
ground state.
For sure, there are also the states where both oscillators are excited. One can be directly convinced that the functions
  \be
\label{sumHerm}
\phi_{nm}(u, z) &=& \sum_{k=0}^m \left( \frac {i\Delta }{4 \sqrt{\Omega_1 \Omega_2}} \right)^k 
\frac {(n-m+k+1)!}{(m-k)! k!} H^+_{n-m+k} H^-_k,\ \ \ \  m\leq n \ , \nonumber \\
\phi_{nm}(u, z) &=& \sum_{k=0}^n \left( \frac {i\Delta }{4 \sqrt{\Omega_1 \Omega_2}} \right)^k 
\frac {(m-n+k+1)!}{(n-k)! k!} H^+_{k} H^-_{m-n+k},\ \ \ \  m >  n  \nonumber \\
&\ &
 \ee
are the eigenfunctions of the operator (\ref{Hhol}) with the eigenvalues (\ref{spec12}). Multiplying the polynomials
(\ref{sumHerm}) by the exponential factors as distated by Eq.(\ref{AnsPsi}), we arrive at the normalizable
wave functions of the hamiltonian (\ref{Ham12}).

We are ready now to see what happens in the limit $\Omega_1 \to \Omega_2$ ($\Delta \to 0$). Two important observations
are in order. 

\begin{itemize}

\item The second exponential factor in (\ref{AnsPsi}) disappears and the wave functions cease to be normalizable. 

\item We see that in the limit $\Delta \to 0$, only the first terms survive in the sums (\ref{sumHerm}) and we obtain
  \be
 \label{limHerm}
 \lim_{\Delta \to 0} \phi_{nm} & \sim & H^+_{n-m} \ \ \ \ \ \ \ \ , \ \ \ \ \ \ m \leq n \nonumber \\
\lim_{\Delta \to 0} \phi_{nm}  & \sim & H^-_{m-n} \ \ \ \ \ \ \ \ , \ \ \ \ \ \ m > n\ . 
  \ee
 In other words, the wave functions depend only on the difference $n-m$, which is { the only} 
relevant quantum number in the limit $\Omega_1 = \Omega_2$.

\end{itemize}

 As this phenomenon is rather unusual and very important for us, let us spend few more words to clarify it. 
Suppose $\Omega_1$ is very close to $\Omega_2$, but still not equal. Then the spectrum includes the sets of nearly 
degenerate states. For example, the states $\Psi_{00}, \Psi_{11}, \Psi_{22}$, etc have the energies $\Delta/2, 3\Delta/2,
5\Delta/2$, etc, which are very close. In the limit $\Delta \to 0$, the energy of all these states coincides, but rather
than having an infinite number of degenerate states, we have only one state: the wave functions 
 $\Psi_{00}, \Psi_{11}, \Psi_{22}$, etc simply {\it coincide} in this limit  by the same token as the eigenvectors
of the matrix  $\left( \begin{array}{cc} 1 & 1 \\ \Delta & 1 \end{array}
\right) $ coincide in the limit $\Delta \to 0$.

\subsection{Interacting theory.}

When $\Omega_1 = \Omega_2$, $u = \bar z$ and the operator (\ref{Hhol}) acquires the form
 \be
\label{Hzzbar}
 \tilde H(z, \bar z) \ =\ \frac 12 \left( \frac \partial {\partial \bar z} -  \frac \partial {\partial  z} \right)^2 +
\Omega \left( \bar z \frac \partial {\partial \bar z} -   z \frac \partial {\partial  z} \right)\ .
 \ee
Its spectrum is bottomless. Let us deform (\ref{Hzzbar}) by
adding there the quartic term $\alpha z^2 \bar z^2$ with positive $\alpha$. Note first of all that it cannot be treated
as a perturbation, however small $\alpha$ is: the wave functions are not normalizable and the matrix elements
of  $\alpha z^2 \bar z^2$ diverge. But one can use the variational approach. Let us take the Ansatz 
  \be
 \label{Ansvar}
 |{\rm var} \rangle \ =\ z^n e^{-Az \bar z} \ ,
 \ee
where $A,n$ are the variational parameters. The matrix element of the unperturbed quadratic hamiltonian
(\ref{Hzzbar}) over the state (\ref{Ansvar}) is
  \be
\label{var2}
\langle {\rm var} | \tilde H | {\rm var} \rangle \ =\ \frac {A(n+1)}2 - \Omega n\ . 
  \ee 
Obviously, by choosing $n$ large enough and $A$ small enough, one can make it as close to $-\infty$ as
one wishes. The bottom is absent and one cannot reach it. For the deformed hamiltonian, the situation is different, however.
We have
 \be
 \label{Evar}
E^{\rm var}(n, A) = \langle {\rm var} | \tilde H + \alpha z^2 \bar z^2| {\rm var} \rangle \ =\ \frac {A(n+1)}2 - \Omega n
+ \frac {\alpha(n+1)(n+2)}{4A^2}\ . 
  \ee
This function has a global minimum. It is reached when 
  \be
A-\Omega - \frac \alpha{4A^2} \ =\  0 
  \ee
and  $n =  {A^3}/\alpha - 2$.

For small $\alpha \ll \Omega^3$,
 \be
A \approx \Omega,\ \ n \approx \frac {\Omega^3}\alpha, \ \ \ \ {\rm and} \ \ \ \ E^{\rm var} \approx -\frac {\Omega^4}{4\alpha}\ .
 \ee
The smaller is $\alpha$, the lower is the variational estimate for the ground state energy and the ground state energy itself.
In the limit $\alpha \to 0$, the spectrum becomes bottomless. 
But for a finite $\alpha$, the bottom exists.
 Note that in the interacting system, the spectrum is completely rearranged compared to the HD oscillator studied above and there is no reason
to expect the peculiar Jordan-like degeneracy anymore. The eigenstates are conventional normalized functions and the solution of the time-dependent 
Schr\"odinger equation has the standard form.

Bearing in mind that $z = \Omega q + ix = \Omega q + i \dot q$, the deformation $\alpha z^2 \bar z^2$ amounts to a 
particular combination of the terms $\sim q^4$, $\sim q^2 \dot q^2$, and $\sim \dot q^4$ in the hamiltonian. For the theory
(\ref{L4mix}) with generic $\alpha, \beta$, the algebra is somewhat more complicated, but the conclusion is the same: in the case 
when the form $\alpha q^4/4 + \beta q^2 x^2/2$ is positive definite, the system has a ground state.

 The requirement of positive definiteness of the deformation is necessary. In the opposite case, choosing the Ansatz 
$$ |{\rm var} \rangle \ \sim  \ (\Omega q + ix)^n \exp\{-Aq^2 - B x^2 \}$$
and playing with $A,B$,   
one can always make the matrix element 
$\langle$ {\sl var} $|$ {\sl deformation} $|$ {\sl var} $\rangle$ negative, 
which would add to the
negative contribution $-\Omega n$ in the variational energy, rather than compensate it. 
The bottom is absent in this case.

\section{Superconformal $6D$ theory}
\setcounter{equation}0

 We start with reminding some basic facts of life for  spinors  in 
$SO(5,1)$ (or rather $Spin(5,1)$) . There are two different complex
4-component spinor representations, the $(1,0)$ spinors $\psi^a$ and the $(0,1)$ spinors $\xi_a$.
In the familiar  $Spin(3,1)$ case, there are   also  
two different spinor representations, which are transformed to each other under complex 
conjugation (on the other hand, complex conjugation leaves an Euclidean $4D$ spinor in the same
representation). 
An essential distinguishing feature of  $Spin(5,1)$ is that complex conjugation
does {\it not} change the type of spinor represenation there (while it does for
Euclidean $6D$ spinors, $Spin(6) \equiv SU(4)$).

Indeed, one can show that the spinor 
   \be
 \bar \psi^a=-C^a_{\dot{a}} \psi^{\dot{a}},
  \ee
is transformed in the same way as $\psi^a$. We defined 
$\psi^{\dot{a}} = (\psi^a)^*$ 
and introduced a symplectic charge-conjugation matrix $C$ satisfying
 \be
C^a_{\dot{a}}C^{\dot{a}}_b=-\delta^a_b\ .
  \ee
The operation $\bar \ $ is the covariant conjugation. 
A somewhat unusual property $\overline{\bar\psi^a}=-\psi^a$ holds. 

 Bearing in mind, however, that $\psi^a$
and $\bar \psi^a$  belong to the same
representation, it is very convenient \cite{HST} to treat them 
on equal footing and 
introduce $\psi^a_{i=1,2} = (\psi^a,  \bar \psi^a)$. The relation
 \be
\label{samosopr}
\bar \psi^a_i =   \psi^{ai} = \epsilon^{ij} \psi_{aj} 
 \ee
holds. 

We choose the  antisymmetric representation of the 6D Weyl matrices
\be
(\gamma^M)_{ab}=-(\gamma^M)_{ba}\,\q\tilde\gamma_M^{ab}=\sfrac12\ve^{abcd}
(\gamma_M)_{cd}
\ee
where $M = 0,1,\ldots,5$ and $\ve^{abcd}$ is the totally antisymmetric symbol.
The basic relations for these Weyl matrices are
\be
&&(\gamma_M)_{ac}(\tilde\gamma_N)^{cb}+(\gamma_N)_{ac}(\tilde\gamma_M)^{cb}
=-2\d^b_a
\eta_{MN},\\
&&\ve_{abcd}=\sfrac12(\gamma^M)_{ab}(\gamma_M)_{cd},
\ee
where $\eta_{MN}$ is the metric of the 6D Minkowski space
($\eta_{00}=-\eta_{11}=\ldots=-\eta_{55}=1)$ and $\gamma_M=\eta_{MN}\gamma^N$.

The generators of the (1,0) spinor representation are $S^{MN} = -\frac12 \sigma^{MN}$,
where
\be
(\sigma^{MN})^b_a=\frac12(\tilde{\gamma}^M\gamma^N- \tilde{\gamma}^N\gamma^M)^b_a,\q
\overline{\sigma^{MN}}=\sigma^{MN}.\lb{sigmadef}\ .
\ee

Supersymmetric field theories are most naturally formulated in the framework of
superspace approach. The $6D$ superspace is more complicated than the 4-dimensional one.
A simple-minded $6D$ superspace involves, besides 6 bosonic coordinates, 8 fermionic
coordinates $\theta^a_i$. However, one can effectively reduce the number of 
fermionic coordinates using the {\it harmonic superspace} approach and working with 
{\it Grassmann analytic} superfields \cite{HSS}. We are not able to dwell on this 
in details and refer the reader to our paper \cite{ISZ}. Here we only present the results.

Let us remind first the form of the conventional quadratic in derivatives
 SYM action in 6 dimensons. It involves the $6D$ gauge field $A_M$, the gluino
field $\psi^a_i$  satisfying (\ref{samosopr}) 
and the triplet of auxiliary fields ${\cal D}_{ik}$. The action reads 
  \be
 \lb{standcomp}
 S = \frac 1{f^2} \int d^6x  \, \mbox{Tr} 
\left \{- \frac 12 F_{MN}^2 - \frac 12 {\cal D}^{ik}  {\cal D}_{ik} + 
i \psi^k \gamma_M \nabla_M \psi_k 
\right\} \ ,
   \ee
where $f$ is the coupling  constant  of canonical dimension  
-1 and $\nabla_M$ is the covariant derivative.

If going down to four dimensions, one reproduces the action 
for ${\cal N} =2$ $4D$  SYM theory. $A_M$ gives the $4D$ gauge field $A_\mu$ and the adjoint
scalar, $\psi^a_i$ gives two $4D$ gluino fields while the triplet of auxiliary fields
can be decomposed into the real auxiliary field $D$ of the 4-dimensional ${\cal N} = 1$ vector multiplet and
the complex auxiliary field $F$ of the adjoint chiral multiplet.

The action of the HD $6D$ gauge theory was derived in \cite{ISZ}.  
 The result is 
\be
 S &=& -\frac{1}{g^2}\int d^6x  \, \mbox{Tr}
\left\{ \left( \nabla^M F_{ML}\right)^2  +  i\psi^j\gamma^M \nabla_M (\nabla)^2\psi_j
 + \frac 12 \left(\nabla_M{\cal D}_{jk}\right)^2
\right. \nn \\
&& \q \left.
+\,  {\cal D}_{lk}{\cal D}^{kj}{\cal D}^{\;\;\;l}_{j}
  -2i {\cal D}_{jk} \left( \psi^j\gamma^M\nabla_M\psi^k
- \nabla_M \psi^j \gamma^M \psi^k \right)
+ (\psi^j\gamma_M \psi_j)^2  \right. \nn \\
&& \q \left. +\, \frac 12  \nabla_M\psi^i
\gamma^M\sigma^{NS}[F_{NS}, \psi_j]
- 2\nabla^M F_{MN}\, \psi^j\gamma^N\psi_j
 \right\}.
\label{CompAct}
 \ee
The lagrangian  
has the canonical dimension 6 and the coupling constant $g$ is dimensionless.

Let us discuss this result. Note first of all  that the quadratic
terms in the lagrangian are obtained from
(\ref{standcomp}) by adding the extra box operator
(it enters with negative sign, this makes the kinetic terms
positive definite in Minkowski space). It is immediately seen for the terms $\propto {\cal D}^2$
and for the fermions. This is true also for the gauge part due to the
identity
 \be
\lb{FboxF}
 {\rm Tr} \left\{ ( \nabla^M F_{MN} )^2 \right \} \
=\ -\frac 12 {\rm Tr} \left\{ F^{MN} \nabla^2 F_{MN} \right\}
-   2i {\rm Tr}\, \left\{ F_{M}^{\;\;\;N} F_{NS} F^{SM} \right\}.
 \ee
The former auxiliary fields ${\cal D}^{ik}$ become dynamical. They carry canonical
dimension 2 and their kinetic
term involves  two derivatives.
There is a cubic term $\propto {\cal D}^3$. This sector of the theory
reminds the renormalizable theory
$(\phi^3)_6$. Gauge and fermion fields have the habitual canonical dimensions
$[A_M] = 1,\ [\psi] = 3/2$. Their kinetic terms involve, correspondingly, 4 and 3 derivatives.
 The lagrangian involves also other interaction terms, all of them having the canonical
dimension 6.

It is instructive to evaluate the number of on--shell degrees of
freedom for this lagrangian. Consider first the gauge field. With the
standard lagrangian $\propto {\rm Tr} \{F_{MN}^2\}$, a
six--dimensional gauge field $A_M$ has  4 on--shell d.o.f. for each
color index. The simplest way to see this is to note that $A_0$
is not dynamical and we have to impose the Gauss law
constraint on the remaining 5 spatial variables.
For the higher-derivative theory, however, the
presence of two extra derivatives doubles the number of d.o.f. and
the correct counting is $2\times 5 = 10$ before imposing the Gauss law constraint
and $10 -1 = 9$ after that. In addition, there are 3
d.o.f. of the fields $D_{ij}$ 
and we have all together 12 bosonic
d.o.f. for each color index. The standard $6D$ Weyl fermion (with
the lagrangian involving only one derivative) has 4 on--shell degrees of
freedom. In our case, we have $4\times 3 = 12$ fermionic d.o.f.
 due to the presence of three derivatives in the kinetic
term. Not unexpectedly, the numbers of bosonic and fermionic degrees of 
freedom on mass shell coincide.

\subsection{Renormalization}
The lagrangian (\ref{CompAct}) does not involve dimensional parameters and is scale--invariant.
 A less trivial and rather remarkable fact
is that the action  is also invariant with respect to special conformal
transformations and the full superconformal group. This is true at the classical level, but,
unfortunately, conformal invariance of this theory is broken by quantum effects.
To see this, let us 
 calculate (at the one--loop level)
the $\beta$ function of our theory. 

\begin{figure}[h]
   \begin{center}
 \includegraphics[width=4.0in]{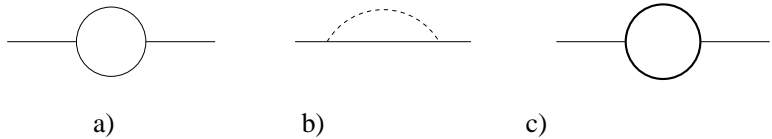}
        \vspace{-2mm}
    \end{center}
\caption{\small Graphs contributing to the renormalization of the kinetic term.
Thin solid lines stand for the particle
${\cal D}$, thick solid lines for fermions, and dashed lines for gauge bosons.}
\label{figD2}
\end{figure}

\begin{figure}[h]
   \begin{center}
 \includegraphics[width=4.0in]{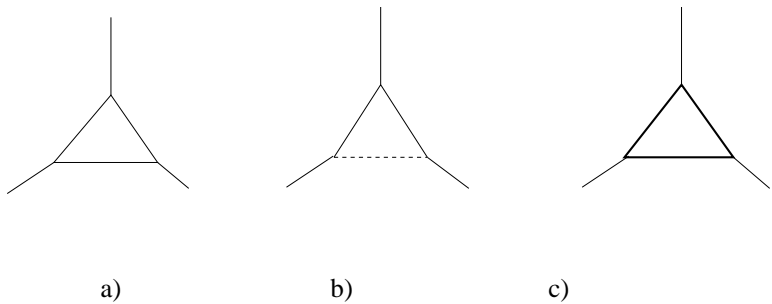}
        \vspace{-2mm}
    \end{center}
\caption{\small The same for the ${\cal D}^3$ vertex.}
\label{figD3}
\end{figure}

The simplest way to do this calculation is to evaluate 1--loop corrections
to the structures $\sim (\pa_M {\cal D})^2$
and $\sim {\cal D}^3$. The relevant Feynman graphs
are depicted in Figs. \ref{figD2}, \ref{figD3}.

For perturbative calculations, we absorb the factor $1/g$
in the definition of the fields. The relevant propagators are
  \be
\lb{prop}
 \langle A_M^A A_N^B \rangle  &=& - \frac {i \eta_{MN} \delta^{AB}}{p^4}\,, \nn \\
\langle \psi^{jA} \psi^{kB} \rangle &=&
- \frac {i \epsilon^{jk} \delta^{AB} p_N \tilde \g^N }{p^4}\,,    \nn \\
 \langle {\cal D}_{ik}^A {\cal D}_{jl}^B \rangle &=&
- \frac {i\delta^{AB}}{p^2} \left( \epsilon_{ij} \epsilon_{kl}
+  \epsilon_{il} \epsilon_{kj} \right),
 \ee
where $A,B$ are color indices, $A_M = A^A_M t^A\,$, etc.
The vertices can be read out directly from the lagrangian.

 Consider first the graphs in Fig. \ref{figD2}.
They involve logarithmic and quadratic divergences.
 The individual quadratically divergent
contributions in the Wilsonean
effective lagrangian are
\be
\lb{quadr}
\Delta {\cal L}^{\rm eff}_{1a} &=& -  \frac {9c_V}2 {\rm Tr}\, \{ {\cal D}_{jk}^2 \} \, I\, , \nn \\
\Delta {\cal L}^{\rm eff}_{1b} &=&   \frac {c_V}2  {\rm Tr}\, \{ {\cal D}_{jk}^2 \} \, I\, ,   \nn \\
\Delta {\cal L}^{\rm eff}_{1c} &=&   4 c_V  {\rm Tr}\, \{ {\cal
D}_{jk}^2 \}\, I\, ,
  \ee
 where $c_V$ is the adjoint Casimir eigenvalue and
 \be
\lb{Iquadr}
 I = \int^\Lambda \frac {d^6p_E}{(2\pi)^6 p_E^4}\,.
 \ee
 We see that the
quadratic divergences cancel out in the sum of the three graphs.
The logarithmic divergences in the 2-point graphs are
  \be
\Delta {\cal L}^{\rm eff}_{(2)} = g^2 c_V \left( - \frac 32 - \frac 76 +
2 \right)  {\rm Tr}\, \{(\pa_M {\cal D}_{jk})^2 \}\, L
= - \frac {2g^2c_V}3  {\rm Tr}\, \{(\pa_M {\cal D}_{jk})^2 \}\, L\,,
  \ee
where
\be
 L = \int_\mu^\Lambda  \frac {d^6p_E}{(2\pi)^6 p_E^6}
= \frac 1{64\pi^3} \ln \frac \Lambda \mu
 \ee
and three terms in the parentheses correspond to the contributions
of the graphs in Fig. \ref{figD2}a,b,c.

The 3-point graphs in Fig. \ref{figD3} involve only logarithmic divergence. We obtain
  \be
\Delta {\cal L}^{\rm eff}_{(3)} = g^3 c_V \left( - \frac 92 - \frac 32 + \frac {32}3  \right)
{\rm Tr}\, \{ {\cal D}_{lk} {\cal D}^{kj} {\cal D}_j^l \}\, L
=  \frac {14 g^3 c_V}3  {\rm Tr}\, \{  {\cal D}_{lk} {\cal D}^{kj} {\cal D}_j^l \}\, L\,.
  \ee
 The full 1-loop effective lagrangian in the ${\cal D}$ sector is
   \be
{\cal L}^{\rm eff}_{{\cal D}} \ =\ - \frac 12
{\rm Tr}\, \{(\pa_M {\cal D}_{jk})^2 \} \left( 1 + \frac {4g^2c_V}3 L \right)
-g  {\rm Tr}\, \{  {\cal D}_{lk} {\cal D}^{kj} {\cal D}_j^l \}
\left( 1 - \frac {14g^2c_V}3 L \right).
 \ee

Absorbing the renormalization factor of the kinetic term
in the field redefinition, we finally obtain
 \be
 \lb{rencharge}
g(\mu) \ =\ g_0 \left( 1 - \frac {20 g_0^2 c_V}3 L \right) = g_0
\left( 1 - \frac {5 g_0^2 c_V} {48 \pi^3} \ln
\frac \Lambda \mu \right)
 \ee
 for the effective charge renormalization.

The sign corresponds to the Landau zero situation, as in the conventional QED.

\section{Discussion}

Our study was motivated by the dream or rather by a sequence of dreams spelled out in the
Introduction. By the reasons outlined there
 \begin{enumerate}
\item We {\it believe} that the TOE is a conventional field theory in multidimensional bulk.
\item We {\it believe} that our Universe represents a thin soap bubble --- 
a classical 3-brane solution in this theory.
\item If the theory claims to be truly fundamental, it should be renormalizable. For $D>4$, 
this means the presence of higher derivatives in the action.
 \item We {\it believe} that for superconformal theories, a way to tackle the HD ghost trouble exists.
 \item We {\it believe} (but not so firmly, this is just the most attractive possibility) that the TOE
enjoys the maximum ${\cal N} = 2$ superconformal symmetry in six dimensions.
\end{enumerate}

Besides dreams, there are also some positive results. First, we constructed a QM HD 
model where the problem
of ghosts {\it is} resolved. Second, 
 we constructed a  nontrivial example of renormalizable
higher-dimensional supersymmetric gauge theory. It is $6D, {\cal
N}{=}(1,0)$ gauge theory with four derivatives in the action and
dimensionless coupling constant. 

Our theory enjoys superconformal
invariance at the classical level, but, unfortunately, the superconformal symmetry 
is anomalous in this case. 
As the
result of this breaking, in accord with the arguments of
\cite{duhi}, the quantum theory suffers from ghosts which can hardly be
  harmless.   

Four-dimensional experience teaches us that though 
nonsupersymmetric, ${\cal N} =1$, and ${\cal N} =2$
supersymmetric theories are anomalous,  the maximum ${\cal N} =4$ supersymmetric 
Yang-Mills theory
is truly conformal --- 
$\beta$ function vanishes there. It is very natural therefore to
{\it believe} that unconstructed yet Holy Grail ${\cal N} =(2,0)$ maximum superconformal $6D$ theory is  
free from anomaly. 

How can it look like ? The first idea coming to mind is to ape the $4D$ construction and to 
couple the $6D$ gauge supermultiplet to $6D$ hypermultiplets. Adding this term to (\ref{CompAct})
one might hope to obtain a theory which would enjoy extended
superconformal symmetry. Unfortunately, this program meets serious technical difficulties and it is not clear at the
moment whether it can be carried out.

The second possibility is that  the ${\cal N} =(2,0)$ theory does not involve at all the gauge
supermultiplet with the action (\ref{CompAct}), but 
depends on tensor rather than vector multiplets \cite{M5,Sok}.
Unfortunately, to describe the tensor multiplet in the framework of
HSS  is not a trivial task which is  not  solved yet. As a result,
no microscopic lagrangian for interacting (2,0) tensor multiplet is
known today...

Finally, one cannot exclude a disapponting possibility 
that the (2,0) theory does not have a lagrangian formulation
whatsoever.  

But the hope dies last !

\end{document}